\documentclass[prb,aps,twocolumn,groupedaddress,floats,showpacs,final]{revtex4}
\usepackage{graphicx}
\usepackage{dcolumn}
\usepackage{bm}
\usepackage{color}
\usepackage{ulem}
\definecolor{blue}{rgb}{0.3,0.3,0.9}

\def\beq{\begin{eqnarray}}
\def\eeq{\end{eqnarray}}
\def\i{{\rm i}}

\begin{document}

\author{Victor Fleurov$^1$ and Anatoly Kuklov$^2$}
\address{$^1$ Raymond and Beverly Sackler Faculty of Exact Sciences, School of Physics and Astronomy, Tel-Aviv University -
Tel-Aviv 69978, Israel }
\address{$^2$ Department of Physics and Astronomy, CSI, and the Graduate Center of CUNY, New York.}


\title{Phases and phase transitions of Bose condensed light}

\date{\today}
\begin{abstract}
Bose-Einstein condensation of light in 2D is characterized by two classical fields corresponding to two polarizations of light as well as by the distribution of dye molecules inducing light thermalization through dipolar transition. In the case when this transition is triple-degenerate the resulting field theory for the  condensate of light is O(4) symmetric, which precludes algebraic long range order in 2D at any finite temperature $T$. If the dipolar degeneracy is removed, then, equilibrium phases with lower symmetries -- O(2)$\times$Z$_2$ and O(2) can emerge. Accordingly, algebraic off diagonal order of light condensate becomes possible.  An orientationsl disorder introduced by local dipolar anisotropy can destroy algebraic order in one-photon density matrix while preserving it in the two-photon one. This represents formation of the condensate of photon pairs.
\end{abstract}


\maketitle

Recent realization of Bose-Einstein condensate of light, Refs.[\onlinecite{Nature,Schmitt,NS14,MN15,GPO18}], opens up new pathways toward exploring properties of light. In these experiments the thermalization of light is achieved due to absorption and reemission of photons by dye molecules which represent a thermal bath -- thanks to their  manifolds of rovibrational states exchanging energy with the solvent [\onlinecite{KK15}]. 
The equilibration of light occurs on much faster scale than the photon escape. Thus, for all practical purposes the emerging phase of light becomes a thermodynamical phase. It is important that, due to a significant difference between temperature and photon energies, photons behave as
conserving particles which can be characterized by finite chemical potential.   

Here we point out that the presence of the dye molecules is not limited to the thermalization of photons. The symmetry of molecules  makes the system very rich in possible phases and phase transitions. Here we take mostly the phenomenological approach 
within Gross-Pitaevskii (GP) functional which takes into account polarization of light as well as the orientation of the dipole moment $\vec{d}$ of the dye molecule transition. Our analysis is based on the assumption that each dye molecule is a two level system (TLS) and all molecules  interact  with light independently from each other. Situations when these TLS interact resonantly with each other, as suggested in Ref.[\onlinecite{Sela}], will be considered elsewhere.

{\em Order parameter}. The order parameter of light is a complex vector $\vec{E}(x,y,z)$ representing the amplitude of electric field in the rotating wave approximation. The geometry of the experiment [\onlinecite{Nature}] selects a single longitudinal mode $\phi(z)$ along the resonator axis Z so that the field can be represented as
 $\vec{E}(x,y,z)=\vec{\psi}(x,y)\phi(z)$, where the complex vector $\vec{\psi}(x,y)=(\psi_x, \psi_y)$ accounts for the transverse
order and its long wave fluctuations in XY-plane.

{\em O(4) symmetry}.
If the dye molecules  were fully symmetric, the dipolar transition would be triple degenerate. Accordingly, each component of $\vec{E}$ would  see no preferential orientation of the molecules, and a contribution to free energy of one molecule exposed to the condensate of light will depend on the product $d |\vec{E}| \sim |\vec{\psi}|$, where $d$ stands for the absolute value of the dipolar moment of the transition.  Accordingly, the potential energy density $U$ of the GP functional can be represented as Landau  expansion [\onlinecite{Landau}] in the modulus $|\vec{\psi}|$ as
%
$U = a |\vec{\psi}|^2 + b |\vec{\psi}|^4$,
%
with some coefficients $a, b>0$. It is important to realize that the resulting GP functional is O(4) symmetric. The phase transition within the Landau approach [\onlinecite{Landau}] corresponds to $a=0$, and the condensation occurs at $a < 0$ with the order parameter modulus defined as $|\vec{\psi}|=\sqrt{-a/2b}$ and the orientation of $\vec{\psi}$  formed spontaneously. 

In 2D, enhanced long range fluctuations of the soft mode, that is, the orientation and phases of the components $\psi_{x,y}$, change the nature of the spontaneous order quite dramatically. For a single complex field $\psi$ characterized by O(2) (or U(1)) symmetry these fluctuations imply that, at best, the emerging order in the density matrix  $\langle \psi^*(\vec{x}) \psi(0)\rangle$ is algebraic, and as temperature increases the Berezinskii-Kosterlitz-Thouless (BKT) transition takes place into a normal phase characterized by exponentially decaying correlations (see in Ref. [\onlinecite{BKT}]) . In the case of O(4) symmetry in 2D there is no phase with algebraic order at any finite temperature [\onlinecite{Shenker}]. The density matrix $\langle \vec{\psi}^*(\vec{x}) \vec{\psi}(0)\rangle$ is exponentially decaying above some correlation length $\xi$. 
 In other words, there is no O(4) symmetric off diagonal condensate in 2D in the thermodynamical limit.

{\em Non-degenerate dipolar transition and the emergence of lower symmetries }.
In most cases the TLS transition is non-degenerate [\onlinecite{Moodie}]. Thus, it is characterized by a local orientation $\vec{d}(x,y,z)$ of its dipole moment. This creates local anisotropy so that the local contribution to free energy can be represented as an expansion in  $|\vec{d}\vec{E}|^2$ -- in a stark  contrast with the triple-degenerate case. The contributions to free energy must reflect such anisotropy. In general, one can write Landau expansion as
\beq
 U=&  \int dxdy\Big[- \, d_{ij}  \psi^*_i\psi_j + d_{ijkl} \psi^*_i \psi^*_j\psi_k\psi_l + \nonumber \\
  &(\eta_i \psi_i^* + c.c.)\Big].
\label{Fd}
\eeq
Here and below the summation runs over the repeated indices corresponding to $x,y$-directions. The tensors $d_{ij}$ and $d_{ijkl}$ can be related to a particular distribution of the TLS molecules and the orientations of their vectors $\vec{d}$ of the transition.
In principle, higher order tensors will also be present in the expansion (\ref{Fd}). It is, however, reasonable to assume that these terms do not qualitatively affect the results presented here.
 There is also a linear in $\psi_i$ contribution to $U$ in Eq.(\ref{Fd}) coming from externally imposed light represented by the amplitude $\vec{\eta}$.

The tensors in Eq.(\ref{Fd}) are introduced after the integrations  $\sim \int dz |\vec{d} \vec{E}|^2  $ and $\sim \int dz |\vec{d} \vec{E}|^4 $ over the width of the resonator along its main axis:
\beq
d_{ij}(x,y) = c_2 \int dz d_i d_j|\phi (z)|^2,
\label{dij}
\eeq
\beq
d_{ijkl}(x,y)= c_4 \int dz d_i d_j d_k d_l|\phi (z)|^4,
\label{dijkl}
\eeq
with some coefficients $c_2>0, c_4>0$ depending on the nature of the density of states of the molecular transition, temperature and density $n_0$ of the TLS molecules. 
To illustrate our approach we consider a single TLS characterized by half of the energy difference $\delta$ between its ground and excited levels in the rotating wave approximation, with $\vec{d}$ being a matrix element of the dipolar transition between the levels along some fixed direction. Then, in the presence of the off-diagonal term $\sim \vec{d}\vec{E}$ the contribution to
the free energy becomes $ F_1=-T\ln\left[\exp(\Delta/T) + \exp(-\Delta/T)\right]$, where $ \Delta \equiv \sqrt{\delta^2+ |\vec{d}\vec{E}|^2}$. The expansion up to the quartic term gives $F_1= - \tilde{a} |\vec{d}\vec{E}|^2 + \tilde{b} |\vec{d}\vec{E}|^4$ , where the following notations are introduced $\tilde{a}= \tanh(\delta/T)/(2\delta)$ and $\tilde{b}=\left(\tanh(\delta/T) - \frac{\delta}{T\cosh^2(\delta/T)}\right)/(8\delta^3)$. [It is possible to see that $\tilde{b} > 0$]. Then, $c_2 \sim \tilde{a} n_0$ and $c_4 \sim \tilde{b}n_0$.

Here we will not be focusing on the microscopic definitions of the coefficients and will conduct a general phenomenological symmetry analysis. 
We consider several cases distinguished by structure of the tensors $d_{ij}$ and $ d_{ijkl}$, and take an approach
where the spatial distribution of $\vec{d}$ is an external frozen variable which provides a disorder to the dynamical field $\vec{\psi}$. Situations where $\vec{d}$ is also a (coarse grained) dynamical  variable will be addressed elsewhere.

\paragraph{\em Emerging isotropy}.
We first consider the case of no dipolar anisotropy  -- that is, no residual coarse grained vector field $\vec{d}_0(\vec{x})$.
Furthermore, we presume that in the absence of any interaction between TLS molecules, the integration along Z-direction 
in Eqs.(\ref{dij},\ref{dijkl}) returns isotropic tensors 
\beq
d_{ij} \to d_2 \delta_{ij},\,\,\, d_{ijkl} \to g\left( \delta_{ij}\delta _{kl} + \delta_{ik}\delta_{jl}  + \delta_{il}\delta_{jk}\right),
\label{iso}
\eeq
with some coefficients, $d_2$ and $g$.
  In other words, there is no disorder introduced into the 2D functional (\ref{Fd}). This assumption is reasonable if along the length of the resonator in Z-direction there are many TLS molecules with random orientations of their microscopic $\vec{d}$.  
  Accordingly, a substitution of Eqs.(\ref{iso}) into Eq.(\ref{Fd}) gives
\beq
U&\equiv& -\mu |\vec{\psi}|^2 + g\left(2|\vec{\psi}|^4 + (\vec{\psi}^*)^2(\vec{\psi})^2\right)-  \nonumber \\
&& (\vec{\eta}\vec{\psi}^*+c.c.).
\label{U}
\eeq
The term $\sim \mu $ corresponds to the effective chemical potential of light. It is determined by the populations factors of the dye molecules and the profiles of the absorption/emission coefficients (see in Ref. [\onlinecite{Schmitt}]). It is important that due to the isotropy the contribution from the term $\sim d_{ij}$ can be absorbed into the definition of the chemical potential $\mu$. Finally, the full GP functional becomes
\beq
H=\int dxdy\Big[i(\vec{\psi}^*\dot{\vec{\psi}}-c.c.) -  \frac{1}{2m} |\nabla_i \vec{\psi}|^2 -U\Big], \label{H}
\eeq
where $m$ stands for the effective mass of photons induced by the longitudinal confinement along the Z-direction [\onlinecite{Nature}].

\begin{figure}[!htb]
	\includegraphics[width=1.0 \columnwidth]{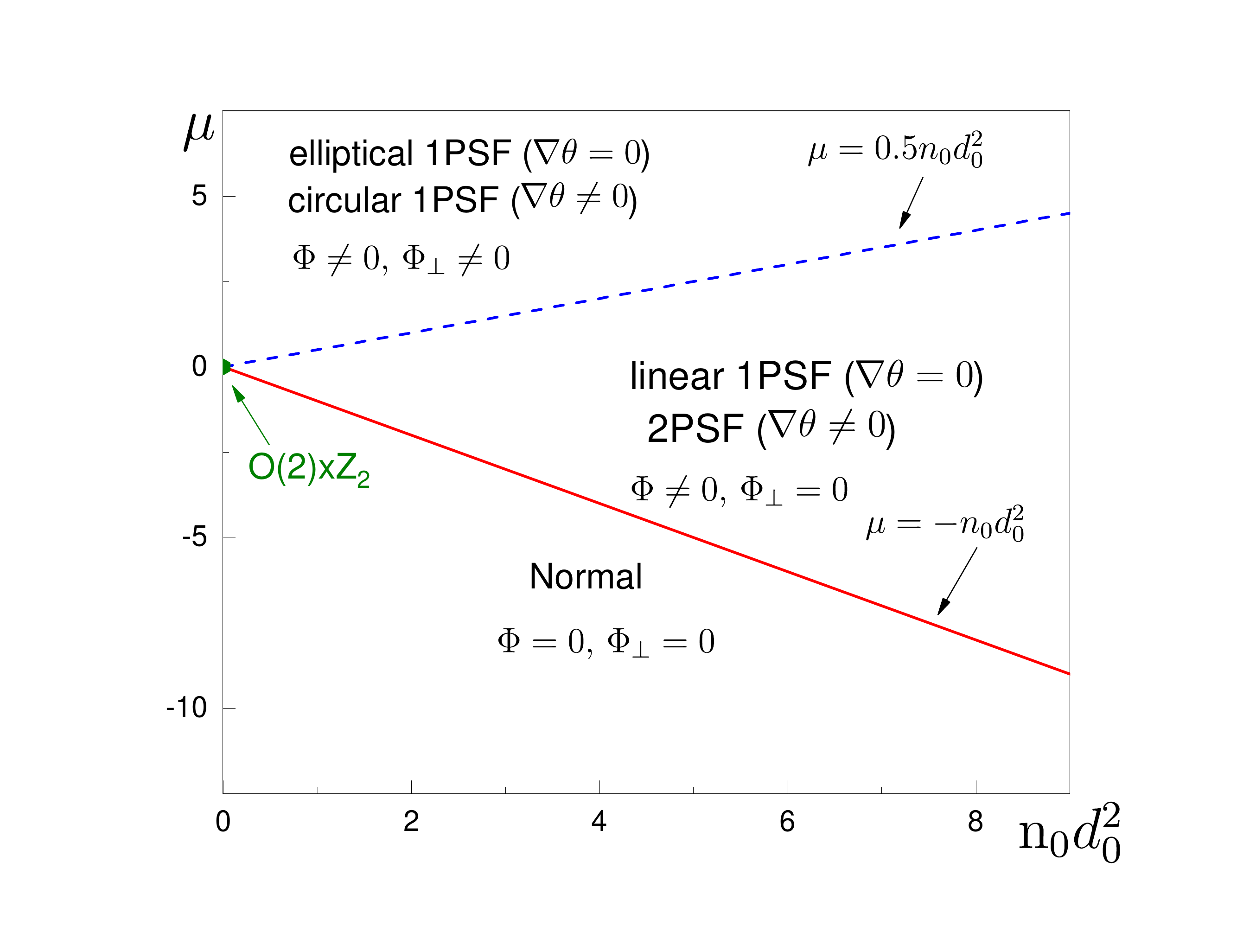}
	\vskip-8mm
	\caption{(Color online) Mean field phase diagram ($\vec{\eta}=0$, units are arbitrary). For the notations see the text. }
	\label{fig1}
\end{figure}

The action (\ref{H},\ref{U}) is characterized by the symmetry O(2)$\times$Z$_2$. Introducing real $\vec{a}$ and imaginary $\vec{b}$ parts of $\vec{\psi}=\vec{a} +i \vec{b}$, the term (\ref{U}) becomes
\beq
U&=& -\mu (\vec{a}^2+\vec{b}^2) +
g\left(3(\vec{a}^2+\vec{b}^2)^2 - 4(\vec{a}\times\vec{b})^2\right) - \nonumber \\
&&\Big((\vec{\eta} + \vec{\eta}^*)\vec{a} + i (\vec{\eta}^*-\vec{\eta})\vec{b}\Big) .
\label{UMF}
\eeq
At the mean field level the condensation corresponds to $\mu > 0$. The lowest energy of the functional (at $\vec{\eta} = 0$) is achieved for $\vec{b} \perp \vec{a}$ and $\vec{a}^2 = \vec{b}^2 = \mu/(8g)$. Explicitly,
 $b_x= a_y$,  $b_y=-a_x$ or $b_x= - a_y$,  $b_y=a_x$ so that
\beq
\psi_x= a_x \pm i a_y,\,\, \psi_y= a_y \mp i a_x,
\label{MF}
\eeq
where the sign $\pm$ is correlated in both equations and it represents two directions of circularly polarized light.
Thus, the ground state of the condensed light is characterized by a spontaneous circular polarization -- left or right handed. This corresponds to  $Z_2$ symmetry. Rotation of the vector $\vec{a}$ (together with $\vec{b}$) in the XY-plane implies O(2) (or U(1)) symmetry. Accordingly, the BEC transition belongs to the O(2)$\times$Z$_2$ universality class, which, in general, differs from the BKT scenario charaterizing O(2) symmetry. On the phase diagram, Fig.~\ref{fig1},
this transition corresponds to the point $\mu=0,n_0d^2_0=0$ labeled as "O(2)$\times$Z$_2$".

In the normal phase, $\mu <0$, the polarization is induced by the external field, $\vec{\psi} \propto \vec{\eta}$.  If $\mu>0$, the polarization of the condensate becomes elliptical because of the competition between the external bias $\vec{\eta}$ and the tendency for spontaneous circular polarization.

The emerging order is algebraic in 2D because of the gaussian rotational fluctuations of the vectors $\vec{a}, \vec{b}$ locked to each other. Thus, the one-photon density matrix (OPDM) $\langle \vec{\psi}^*(\vec{x})\vec{\psi}(0) \rangle \sim 1/|\vec{x}|^K$ with some index $K>0$ determined by the parameters and temperature (see in Ref.[\onlinecite{BKT}]). Such a dramatic change from the O(4) symmetric case, where the density matrix is exponentially decaying, is due to the anomalous term $\sim (\vec{\psi}^*)^2(\vec{\psi})^2$ in the energy (\ref{U}). This term lowers the symmetry from O(4) to O(2)$\times $ Z$_2$ and, thus, makes the condensation possible. It should also be noted that the product $S_z=\vec{a}\times \vec{b} \sim i\vec{\psi}^* \times \vec{\psi}$ represents the order parameter for the Z$_2$ order, which is {\it long} ranged -- similarly to 2D Ising model for the spin variable $S_z \sim \pm 1$ .

An interesting question is about the nature of the O(2)$\times$Z$_2$ transition. There are two options: i) It is a single  transition -- continuous or discontinous; ii) The condensation can proceed as two successive transitions corresponding to breaking the full symmetry one after the other -- first Z$_2$ becomes broken and, then, O(2) follows as temperature lowers further down. If this occurs, the variable $S_z$ becomes condensed while vectors $\vec{a}$ and $\vec{b}$ remain disordered (through their joint rotations in the XY plane). This situation would correspond to the formation of a {\it two-photon} condensate. These options will be analyzed elsewhere. Here and below we will be using the abbreviation mPSF (m-photon superfluid) to characterize m-photon (m=1,2) algebraic order.

We note that the field theories with U(1)$\times$Z$_2$ symmetry  describe multi-band superconductors with spontaneous time reversal symmetry breaking
[\onlinecite{Babaev_2011,Chubukov,Babaev_2013}]. These feature a variety of topological excitations  [\onlinecite{Babaev_2011}]
and also allow for the splitting of the transition into two [\onlinecite{Babaev_2013}].
A significant difference between these systems and the condensate of light lies in the structure of order parameters -- scalars in Refs.[\onlinecite{Babaev_2011,Chubukov,Babaev_2013}] and a complex vector $\vec{\psi}$ for the light. This, in particular, determines the effect which is not present in superconductors -- paired photonic condensate induced by non-Abelian gauge disorder. This effect is described below.

\paragraph{\em Double degenerate transitions}. Here we comment on the situation of a double-degenerate TLS transition. 
If the plane of the dipolar degeneracy coincides with the XY plane of the resonator, then the contribution from each TLS is determined by $\sim |\vec{\psi}|$, which results in the O(4) symmetry. More realistic situation is when the molecular plane of each TLS is randomly oriented with respect to the resonator. Then, the contribution from each TLS can be represented as powers of $ d|\vec{n}_d \times \vec{E}|$ where $\vec{n}_d=(n_x,n_y,n_z)$ stands for the normal to the degeneracy plane. Then, averaging of the terms $\sim |\vec{n}_d \times \vec{E}|^2$ and $ \sim |\vec{n}_d \times \vec{E}|^4$ over all the possible 3D orientations of $\vec{n}_d$ will result in the form similar to (\ref{U}) with the anomalous term $\sim \vec{\psi}^{*2} \vec{\psi}^{2}$ present.  Thus, this case is qualitatively the same as the non-degenerate situation.

\paragraph{\em Dipolar anisotropy}.
Let's consider a situation when there is a residual dipolar anistropy $\vec{d}_0(x,y)$, and begin with the case when this vector is a constant along the XY plane. Then, in addition to the isotropic contributions, there is a term $ - n_0| \vec{d}_0 \vec{\psi}|^2$ in the functional (\ref{U})
favoring co-linear orientation of $\vec{\psi}$ and $\vec{d}_0$. [There is also the term $\sim |\vec{d}_0\vec{\psi}|^4$ which can be ignored in the limit $|\vec{d}_0| \to 0$].  Accordingly, the  condensation of such a component will occur first. This changes the nature of the condensation -- it proceeds as O(2) transition according to the BKT scenario (see in Ref.[\onlinecite{BKT}]) into the phase with $\vec{\psi} \propto \vec{d}_0$. Thus, the condensate polarization is linear and determined by the vector $\vec{d}_0$ (if ignoring $\vec{\eta}$). 

In general,  $\vec{d}_0$ and the order parameter $\vec{\psi}$ can be represented as
\beq
\vec{\psi} = \vec{n} \Phi + {\rm i} \vec{n}_\perp\Phi_\perp,\quad  \vec{d}_0= d_0 \vec{n}, \quad \vec{n}=(\cos\theta, \sin\theta) ,
\label{Phi}
\eeq
where $\vec{n}$, $\vec{n}_\perp=(\sin\theta, -\cos\theta)$ are unit vectors along and perpendicular to $\vec{d}_0$, respectively, and $\Phi$, $\Phi_\perp$ are complex fields.
Using this representation in Eq.(\ref{H},\ref{U}) with the additional term $ - n_0| \vec{d}_0 \vec{\psi}|^2= -n_0 d_0^2|\Phi|^2$ ,
the GP functional (\ref{H},\ref{U}) becomes
\begin{widetext}
\beq
H&=&\int dxdy\left[i(\Phi^*\dot{\Phi}-c.c.)+i(\Phi_\perp^*\dot{\Phi}_\perp-c.c.) -  \frac{1}{2m}\left[ |\vec{\nabla} \Phi  +i (\vec{\nabla} \theta) \Phi_\perp|^2 + 
|\vec{\nabla} \Phi_\perp  +i (\vec{\nabla}\theta) \Phi|^2\right]
 -U\right], \label{HH} \\
U&=& - (\mu + n_0 d_0^{2}) |\Phi|^2 - \mu |\Phi_\perp|^2 +
g\left[3(|\Phi|^4 + |\Phi_\perp|^4) +4|\Phi|^2|\Phi_\perp|^2 - (\Phi^{*2}\Phi^{2}_\perp +c.c. )\right] - \nonumber \\
&&[\vec{\eta}^*(\vec{n}\Phi + {\rm i}\vec{n}_\perp \Phi_\perp) + c.c.].
\label{grad}
\eeq
\end{widetext}
It is interesting to note that $\vec{\nabla}\theta $ plays here a role of an external (non-Abelian) SU(2) gauge field.
 Since the  chemical potential $\mu_\Phi=\tilde{\mu} +n_0d_0^2$ for the field $\Phi$ is higher than that, $\tilde{\mu}\equiv \mu - (\vec{\nabla}\theta)^2/(2m)$, for $\Phi_\perp$, the field $\Phi$ condenses first. This transition is shown by the solid line $\mu =- n_0d^2_0$ in Fig.~\ref{fig1} (see Appendix \ref{A} for details). The resulting phase is a linearly polarized 1PSF of $\Phi$ marked there as "linear 1PSF $(\vec{\nabla}\theta=0)$". 
The second transition of Z$_2$ universality occurs along the dashed line, Fig.~\ref{fig1}, determined by the condition $\mu=0.5 n_0d_0^2$ (see Appendix \ref{A} for details). This corresponds to an emergence of  elliptical polarization (marked as "elliptical 1PSF $(\vec{\nabla}\theta=0)$" in Fig.~\ref{fig1}).

\paragraph{\em Two-photon condensate}.
Now let's consider a situation when $\vec{d}_0$ is not a constant. We start with the case $|\vec{d}_0 |=$const, so that the disorder is introduced through the gauge  $\vec{\nabla}\theta$ into Eq.(\ref{HH}).  The field $\vec{d}$ is truly disordered above some scale $\xi_d$ if  $\vec{\nabla}\theta(\vec{x})$ is characterized by windings typical for a plasma of proliferated (frozen) vortex-antivortex pairs at a finite density $\sim 1/\xi_d^2$.
 
We note that the effect of $\vec{\nabla}\theta$ becomes essential only in the presence of both components $\Phi,\, \Phi_\perp$. Thus, above the solid line and below the dashed line in Fig.~\ref{fig1}, where $\Phi_\perp=0$ and $\Phi\neq 0$,  
 $\nabla \theta$ can be ignored as long as its correlation length $\xi_d$ becomes larger than the correlation length $\xi \approx 1/\sqrt{m(\mu + n_0 \vec{d}_0^{\,\,2})}$  of $\Phi$ ((see Appendix \ref{A} for more detail). 
Accordingly, the functional (\ref{HH},\ref{grad}) as well as the corresponding correlator $\langle \Phi^*(\vec{x})\Phi(0)\rangle$  decouple from the disorder.
Furthermore, according to the Harris criterion (see in Ref.[\onlinecite{BK}]), the disorder produced by  $ \sim (\nabla \theta)^2$ in Eq.(\ref{HH}) is diagonal (where $\Phi_\perp=0$) and, thus, is irrelevant at the transition marked by the solid line in Fig.~\ref{fig1}. This means that the transition remains of the BKT type even in the
presence of the disordered field $ \sim (\nabla \theta)^2$. 

It is important to note that, despite the emerging order
in the field $\Phi$, the OPDM %
\beq
\langle \psi^*_i(\vec{x}) \psi_j(\vec{x}')\rangle = n_i(\vec{x}) n_j(\vec{x}')  \langle \Phi^*(\vec{x}) \Phi(\vec{x}')\rangle 
\label{one}
\eeq
in the representation (\ref{Phi}) is {\it disordered}. 
Indeed, the factor $n_i(\vec{x}) n_j(\vec{x}')$ in Eq.(\ref{one}) washes out any remnants of the order in $\Phi$ once
the distance $|\vec{x}-\vec{x}'|$ exceeds $\xi_d$. Thus,  there is no 1PSF even though the correlation function $\langle \Phi^*(\vec{x}) \Phi(0)\rangle$ is algebraic. 
Since $\vec{n}$ and $\Phi$ are not, practically, coupled, the averaging over the disorder $\langle ... \rangle_{\theta}$ in Eq.(\ref{one}) can be applied to
the factor  $ n_i(\vec{x}) n_j(\vec{x}')$ only. This produces exponential behavior $\langle n_i(\vec{x}) n_j(\vec{x}')\rangle_{\theta} \sim \exp(- |\vec{x}- \vec{x}'|/\xi_d)$ and, accordingly, $ \langle \psi^*_i(\vec{x}) \psi_j(\vec{x}')\rangle \sim \exp(- |\vec{x}- \vec{x}'|/\xi_d)$, where the algebraic factor is ignored.

The situation is completely different for the two-photon density matrix $\rho^{(2)}_{ijkl}(\vec{x},\vec{x}')=\langle \psi^*_i(\vec{x}) \psi^*_j(\vec{x}) \psi_k(\vec{x}') \psi_l(\vec{x}') \rangle $ -- it does feature the algebraic order. This is a direct consequence of the relation 
 $ \vec{\psi}^{\,\,2}= \Phi^2$, where no vector $\vec{n}$ is present.
In terms of the components:
\beq
\rho^{(2)}_{ijkl}= \langle n_i(\vec{x}) n_j(\vec{x}) n_k(\vec{x}') n_l(\vec{x}')\rangle_\theta  \langle (\Phi^*(\vec{x}))^2 (\Phi(\vec{x}'))^2\rangle.
\label{two}
\eeq
%
At the distances $|\vec{x}-\vec{x}'|<\xi_d$ the factor $\langle n_i n_j n_k n_l\rangle_\theta$ can be represented as $ \approx (\delta_{ij} \delta_{kl} + \delta_{ik}\delta_{jl} + \delta_{il}\delta_{jk})/8$ while at larger distances it $\to \delta_{ij} \delta_{kl}/4$.  Overall, however, the two-photon density matrix is long ranged $\rho^{(2)}_{ijkl} \propto \langle (\Phi^*(\vec{x}))^2 (\Phi(\vec{x}'))^2\rangle$. More specifically,  the component $ \rho^{(2)}_{xxxx}$ changes from
$(3/8)  \langle (\Phi^*(\vec{x}))^2 (\Phi(\vec{x}'))^2\rangle$ at short distances to $(1/4)  \langle (\Phi^*(\vec{x}))^2 (\Phi(\vec{x}'))^2\rangle$ at large distances.
For the component $\rho^{(2)}_{xxyy}$ the corresponding limits are $(1/8)  \langle (\Phi^*(\vec{x}))^2 (\Phi(\vec{x}'))^2\rangle$ and $(1/4)  \langle (\Phi^*(\vec{x}))^2 (\Phi(\vec{x}'))^2\rangle$. Finally, for the component $\rho^{(2)}_{xyxy}$ the limits are $(1/8)  \langle (\Phi^*(\vec{x}))^2 (\Phi(\vec{x}'))^2\rangle$ and $\to 0$. This behavior implies  the emergence of 2PSF while there is no 1PSF.
This phase is labeled as " 2PSF $(\vec{\nabla}\theta\neq 0)$" in Fig.~\ref{fig1}.
 We will analyze the detection schemes of 2PSF based on the two-photon polarization properties elsewhere. 

The condensation of the field $\Phi_\perp$ can restore the one-photon order by introducing vortices into the fields $\Phi,\, \Phi_\perp$ which
compensate partially the disorder created by $\theta(\vec{x})$.
This mechanism corresponds to the transformation of the phase gradient $\vec{\nabla}\varphi$ of the fields $\Phi$ and $\Phi_\perp$ from being vortex free to acquiring windings as  $\vec{\nabla}\varphi \to  \vec{\nabla}\varphi = \vec{\nabla}\tilde{\varphi} - \vec{\nabla}\theta$, with $\tilde{\varphi}$ denoting the irrotational part of the phase.
 The mean field analysis of such a transformation is explained in Appendix \ref{A}. As a result, the OPDM acquires algebraic order of circularly polarized light. This phase occurs above the dashed line in Fig.~\ref{fig1} and it is labeled as "circular 1PSF ($\vec{\nabla}\theta \neq 0$)". In this phase the gauge disorder due to $\vec{\nabla}\theta \neq 0$ becomes irrelevant
and the field $\vec{\psi}$ acquires the coherent components
\beq
\vec{\psi} \sim  (1, \pm \i) \exp(\i \tilde{\varphi}), 
\label{coh}
\eeq
describing circularly polarized 1PSF condensate.
More details can be found in the Appendix \ref{A}.
The nature of the transformation from 2PSF to such a 1PSF requires a separate analysis.  

Finally, we note that including a weak disorder in $d_0$ in Eq.(\ref{grad}) should not modify the above conclusions -- at least for the O(2) transition (see in Ref.[\onlinecite{BK}]). However, once the disorder becomes strong, a Bose glass phase can emerge.

{\em In summary}, the phenomenological analysis of condensed phases of light points out to multiple possibilities characterized by various symmetries and symmetry breaking patterns strongly dependent on the nature of the TLS systems. The algebraic condensation of light in 2D can only occur if the TLS transition 
is not fully degenerate. Then, the resulting symmetry of the condensate is O(2)$\times$Z$_2$, and the algebraic long range order becomes possible. Orientational anisotropy in the TLS ensemble can break this transition into two. The disorder induced by the anisotropy can destroy long-range correlations in the one-photon density matrix, while retaining algebraic order in the two-photon density matrix. Methods for detecting multi-photon condensates need to be developed.

 {\it Acknowledgments}.
We acknowledge useful discussions with Vladimir Yurovsky and Egor Babaev.
This work was supported by the National Science Foundation under the grant DMR1720251.

\appendix 
\section{Mean Field analysis}\label{A}
If the field $\Phi_\perp =0$, the GP functional (\ref{HH},\ref{grad}) becomes (at $\vec{\eta}=0$):
\begin{widetext}
\beq
H=\int dxdy\left[i(\Phi^*\dot{\Phi}-c.c.) -  \frac{1}{2m}|\vec{\nabla} \Phi|^2 +  \left(\frac{ (\vec{\nabla}\theta)^2}{2m} -\mu - n_0d^2_0\right)  |\Phi|^2 - 3g|\Phi|^4 \right].
\label{HHsm}
\eeq
\end{widetext}
Thus, the gauge effect of the field $\vec{\nabla}\theta$ vanishes. The remaining term $\sim (\vec{\nabla}\theta)^2$ contributes to the suppression of local chemical potential. In other words, this term introduces diagonal disorder.  However, in the limit under consideration the contribution $\sim (\nabla \theta)^2$ can be ignored if compared with $\mu + n_0d^2_0$. This implies that $\Phi$, practically, decouples from the external gauge field $\vec{\nabla}\theta$ and forms algebraic order at $\mu + n_0d^2_0 - \langle (\vec{\nabla}\theta)^2/(2m)\rangle_\theta \approx \mu + n_0d^2_0 >0$. In contrast, the total field $\vec{\psi}= \vec{n} \Phi $, Eq.(\ref{Phi}), shows no such an order
because $\vec{n}=(\cos\theta,\sin\theta)$ winds in space and washes out any coherence contributed by $\Phi$. This is the mechanism which destroys 1PSF and induces 2PSF while crossing the solid line from the normal phase in Fig.1. The 2PSF can be characterized by the order parameter $\vec{\psi}\vec{\psi} = \Phi^2$, which exhibits algebraic order consistent with the functional (\ref{HHsm}).

In the case when both fields are condensed, the mean field solution for (\ref{HH},\ref{grad}) gives
\beq
\Phi=\Phi_1 \exp(\i \varphi),\, \Phi_\perp=\Phi_2\exp(\i \varphi),\,\, \nonumber \\
\Phi^2_1=\frac{2\mu + 3n_0d^2_0}{16g},\,\, \Phi_2^2=\frac{2\mu - n_0d^2_0}{16g}>0,
\label{MF}
\eeq
which is valid for $\mu > 0.5n_0d^2_0$.
Here the phase $\varphi$ is to be determined from the minimum of the gradient part
\begin{widetext}
\beq
\Delta H= \int dxdy \frac{1}{2m}\left[(\Phi_1 \vec{\nabla} \varphi + \Phi_2 \vec{\theta})^2 + (\Phi_2 \vec{\nabla} \varphi + \Phi_1 \vec{\theta})^2\right]
\label{grrr}
\eeq
\end{widetext}
of the functional (\ref{HH}).

Let's assume the field $ \vec{\nabla} \theta $ has just a single vortex. Then, the minimum of $\Delta H$ can be reached either for $\vec{\nabla}\varphi=0$
(which corresponds to 2PSF) or for $\vec{\nabla}\varphi =- \vec{\nabla} \theta$, which (as will be explained below) corresponds to circulary polarized 1PSF. Comparison of these two energies leads to the condition
\beq
\frac{\Phi^2_2}{\Phi^2_1} < 7-4\sqrt{3}
\label{cond}
\eeq
for the existence of the 2PSF. Using the solution (\ref{MF}) in Eq.(\ref{cond}) gives
\beq
\mu < \frac{3\sqrt{3} -5}{2\sqrt{3}-3} n_0d^2_0 \approx 0.423 n_0 d_0^2,
\eeq
which conflicts with the requirement $\mu > 0.5n_0d^2_0$ needed to have $\Phi^2_2 >0$.

Thus, for  $\mu > 0.5n_0d^2_0$ the minimum of $\Delta H$ is achieved for $\vec{\nabla}\varphi = - \vec{\nabla} \theta$, that is when the phase
of the fields $\Phi, \Phi_\perp$ contains antivortex. In this case the components of the full field $\vec{\psi}$ become
\beq
\psi_x &=& \left[\frac{\Phi_1 + \Phi_2}{2} + \frac{\Phi_1 - \Phi_2}{2}\exp(2\i \theta)\right] \exp(\i \tilde{\varphi}), \\
\psi_y &=& -\i \left[\frac{\Phi_1 + \Phi_2}{2} + \frac{\Phi_2 - \Phi_1}{2}\exp(2\i \theta)\right] \exp(\i \tilde{\varphi}),
\eeq
where $\tilde{\varphi}$ is a non-winding part of $\varphi$, that is, which accounts for non-topological excitations --  phonons of photonic condensate.
The terms $\sim \exp(2\i \theta)$ containing the winding field are washed out at large
distances, and the coherent part of the total field $\vec{\psi}$ becomes
\beq
\psi_x &=& \frac{1}{2}(\Phi_1 + \Phi_2) \exp(\i \tilde{\varphi}), \\
\psi_y &=& - \frac{\i}{2}(\Phi_1 + \Phi_2) \exp(\i \tilde{\varphi}),
\eeq
which describes 1PSF with circular polarization as depicted in Eq.(\ref{coh}).
Such a transformation renders the gauge field $\vec{\nabla}\theta$ irrelevant, and the algebraic part of the OPDM becomes
\beq
\sim  \langle \exp( i [\tilde{\varphi}(\vec{x}) - \tilde{\varphi}(\vec{x}')]) \rangle ,
\eeq
 which after the gaussian averaging over phonons produces algebraic decay at large distances (see in Ref.[\onlinecite{BKT}]).
Thus, for $\mu >0.5n_0d^2_0$, the resulting phase is 1PSF and it is characterized by circular polarization. According to the mean field analysis
it occurs as soon as both fields are condensed -- that is, above the dotted line in Fig.1.

\end{document}